\documentclass[traditabstract]{aa} % for the abstract without structuration 
\usepackage{graphicx}
\usepackage{txfonts}
\usepackage{natbib}
\usepackage{url}
\begin{document}
   \title{Resolving the molecular gas around the lensed quasar \object{RXJ0911.4+0551}\thanks{Based on observations carried out with the IRAM Plateau de Bure Interferometer. IRAM is supported by INSU/CNRS (France), MPG (Germany) and IGN (Spain).}}

   %\subtitle{}
   	\author{
   	P. T. Anh \inst{1,2,3,4}
    \and F. Boone \inst{1,2}
    %\and Pierre Darriulat  \inst{1} 
    \and D. T. Hoai \inst{3,4} 
    \and P. T. Nhung \inst{3} 
    \and A. Wei\ss \inst{5}	
    \and J.-P. Kneib \inst{6}
    \and A. Beelen \inst{7}
    \and P. Salom\'e \inst{8}
     }
          
   \institute{
     Universit\'e de Toulouse; UPS-OMP; IRAP; Toulouse, France
     \and
     CNRS; IRAP; 9 Av. colonel Roche, BP 44346, F-31028 Toulouse cedex 4, France
     \and
     VATLY, INST, 179 Hoang Quoc Viet, Cau Giay, Hanoi, Vietnam
     \and
     Institute of Physics, 10 Dao Tan, Hanoi, Vietnam
     \and
     Max-Planck Institut f\"ur Radioastronomie, Auf dem H\"ugel 69, 53121 Bonn, Germany
     \and 
     Laboratoire d'Astrophysique de Marseille, CNRS-Universit\'e Aix-Marseille
     \and 
     Institut d'Astrophysique Spatiale, B\^at. 121, Universit\'e Paris-Sud, 91405 Orsay Cedex, France
         \and
     LERMA, Observatoire de Paris, UMR 8112 du CNRS, 75014, Paris, France
   }
  
   \date{Received ; accepted}

% \abstract{}{}{}{}{} 
% 5 {} token are mandatory
 
  \abstract{
  We report on high angular resolution observations of the CO(7--6) line and millimeter continuum in the host galaxy of the gravitationally lensed (z$\sim$2.8) quasar \object{RXJ0911.4+0551} using the Plateau de Bure Interferometer. Our CO observations resolve the molecular disk of the source. Using a lens model based on HST observations, we fit source models to the observed visibilities. We estimate a molecular disk radius of 1$\pm$0.2\,kpc and an inclination of 69$\pm$6\,$\deg$, the continuum is more compact and is only marginally resolved by our observations. The relatively low molecular gas mass, $M_{\rm gas}=(2.3\pm 0.5)\times 10^{9}$\,M$_{\odot}$, and far-infrared (FIR) luminosity, $L_{\rm FIR}=(7.2\pm 1.5) \times 10^{11}$\,L$_{\odot}$, of this quasar could be explained by its relatively low dynamical mass, $M_{\rm dyn}=(3.9\pm 0.9)\times 10^9$\,M$_{\odot}$. It would be a scaled-down version the quasi-stellar objects (QSOs) usually found at high-$z$. The  FIR and CO luminosities lie on the correlation found for QSOs from low to high redshifts and the gas-to-dust ratio ($45\pm 17$) is similar to the one measured in the z=6.4 QSO, { \object{SDSS\,J1148+5251}}. Differential magnification affects the continuum-to-line luminosity ratio, the line profile, and possibly the spectral energy distribution. 
 }


   \maketitle
%
%________________________________________________________________
%
\section{Introduction}

In the past 15 years, submillimeter/millimeter (submm) observations of the interstellar medium in high-redshift (high-z) galaxies have opened a new window for studying the early Universe and the evolution of galaxies. In particular, continuum emission from dust heated by intense star formation ($\ge 100$\,M$_{\odot}$\,yr$^{-1}$)  as well as large amounts of molecular gas ($\ge 10^{10}$\,M$_{\odot}$) have been observed in  quasi-stellar objects (QSOs) back to the end of the epoch of reionization \citep{2005ARA&A..43..677S,2013arXiv1301.0371C}. 
Simultaneously, a population of high-z galaxies  dubbed submillimeter galaxies \citep[SMGs,][]{2002PhR...369..111B,2011MNRAS.412.1913I,2011ApJ...739L..31R} with similar molecular gas masses, star formation rates (SFR) and evolution have been discovered. Recent observations suggest  a possible evolutionary  link between these two populations.  The large gas reservoirs observed in SMGs are supposed to be driven to the galaxy center, feeding intense starbursts and leading to the growth and activity of the central black holes that would become QSOs. The SMGs would thus transform into QSOs on a time-scale of $\sim$100\,Myr \citep[][]{2006AJ....131.2763C, 2008MNRAS.389...45C,2012MNRAS.426.3201S}.  If confirmed, this scenario would link the history of cosmic star formation with that of black hole growth, and it would give new insights into the role of feedback mechanisms in galaxy evolution. However, more detailed studies of individual galaxies are required to verify whether the morphologies and star-forming properties are consistent with this scenario.

Gravitational lensing is an essential tool for conducting such studies at sensitivities and physical resolutions that would otherwise be out of reach.  The spatial stretching  induced by lensing combined with high-resolution instruments allows us to obtain information on the morphology and the kinematics of the source. The pioneer observation of the Cloverleaf using the CO(7--6) line \citep{1997A&A...321...24A}
was the first to demonstrate the power of such studies. Resolving the source is also important for correctly interpreting the integrated fluxes of  lensed systems because the regions observed in line and continuum  or at different frequencies may not overlap and may therefore undergo differential magnification.

In this paper we present high-resolution Plateau de Bure Interferometer (PdBI) observations of the CO(7--6) line and the millimeter continuum emission in the quadruply lensed quasar \object{RXJ0911.4+0551} at $z=2.796$ \citep{1995A&AS..110..469B,1995A&AS..111..195H,1997A&A...317L..13B,2012ApJ...753..102W}. 
This QSO is particularly interesting because the lens, which includes the contribution of an intervening  galaxy cluster, has previously been well-studied \citep{1998ApJ...501L...5B, 2000ApJ...544L..35K}, and recent observations revealed a molecular gas mass $M_{\rm gas}=4.2\times10^{9}$\,M$_{\odot}$ \citep{2004ApJ...609...61H} and a far-infrared (FIR) luminosity $L_{\rm FIR}=0.8\times10^{12}$\,L$_{\odot}$ \citep{2009ApJ...707..988W} that are an order of magnitude lower than usually observed in QSOs. By measuring the source size and morphology we aim at measuring the dynamical mass to find out whether these low values are measured because this is a lower mass galaxy or because this is an evolved galaxy that has consumed its gas already. We also investigate the effects of differential magnification on the continuum/line luminosity ratio and on the spectra. { We adopt the $\Lambda$CDM concordance cosmology: $H_0=71$\,km\,s$^{-1}$\,Mpc$^{-1}$, $\Omega_M=0.27$ and $\Omega_{\Lambda}=0.73$.}

\section{Observations and results}

\begin{figure*}[htbf]
   \centering
   \resizebox{15cm}{!}{
      \includegraphics[height=5cm]{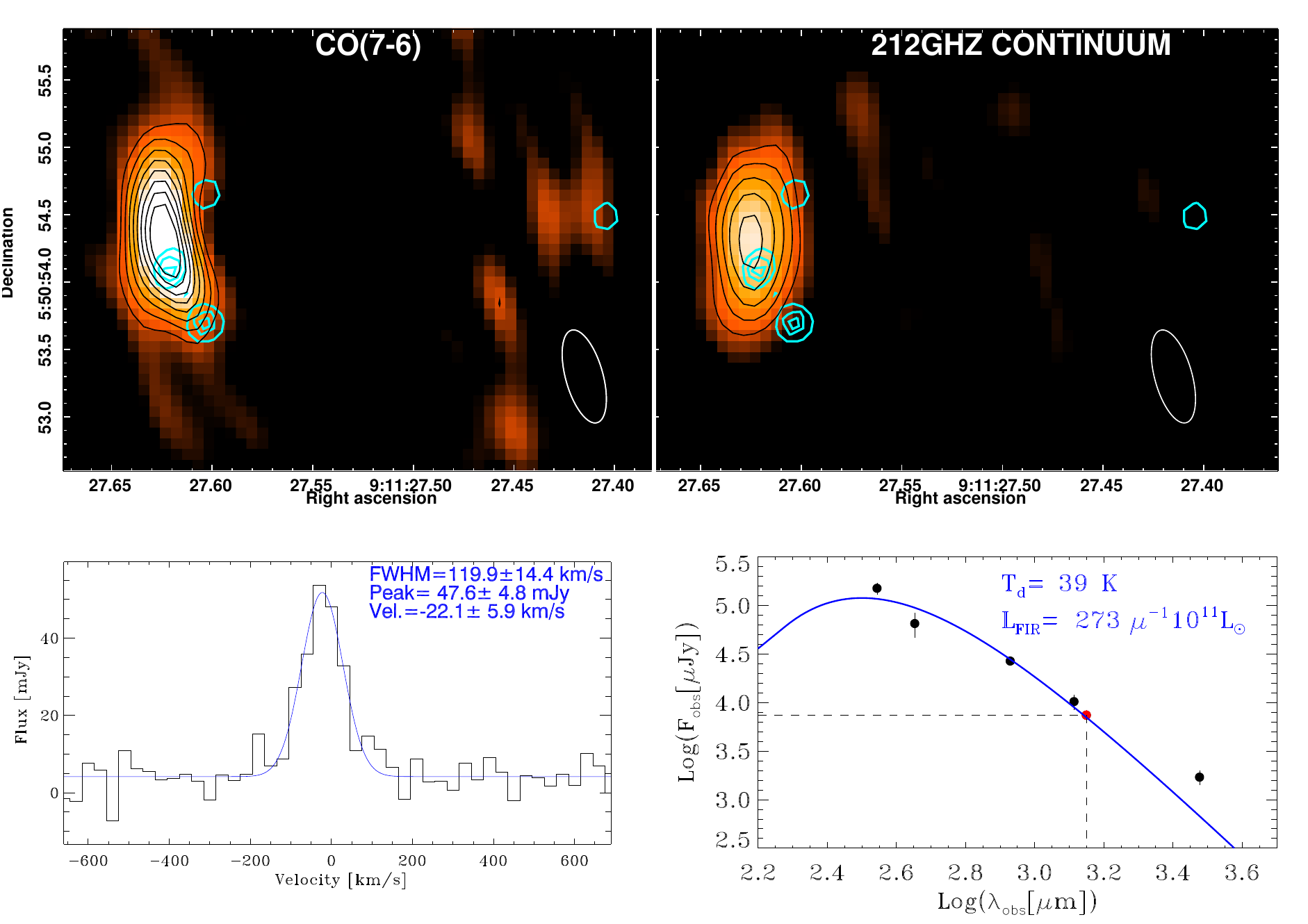}
   }

   \caption{{\it Upper left:} CO(7--6) emission integrated over the velocity range $[-120, 60]$ km\,s$^{-1}$. The black contour levels start at 3$\times\sigma$ and are spaced by $\sigma=135$\,mJy\,beam$^{-1}$\,km\,s$^{-1}$. The white ellipse at the bottom right represents the beam at half maximum (0.71$''\times$0.28$''$). The HST map contours are overlaid in cyan. {\it Upper right:} The 212\,GHz continuum map. The black contour levels start at 3$\times\sigma$ and are spaced by $\sigma=0.28$\,mJy\,beam$^{-1}$. {\it Lower left:} The CO(7--6) integrated spectrum with the best Gaussian fit overlaid. {\it Lower right:} best-fit modified black body spectral energy distribution to the FIR continuum measurements derived by \citet{2009ApJ...707..988W}, \citet{2002ApJ...571..712B}, and us (red dot at $\lambda=$1.41\,mm).}
   \label{fig:obs}%
\end{figure*}

\begin{figure}
   \centering
  \setlength{\unitlength}{\textwidth}
   \resizebox{8cm}{!}{
     \includegraphics{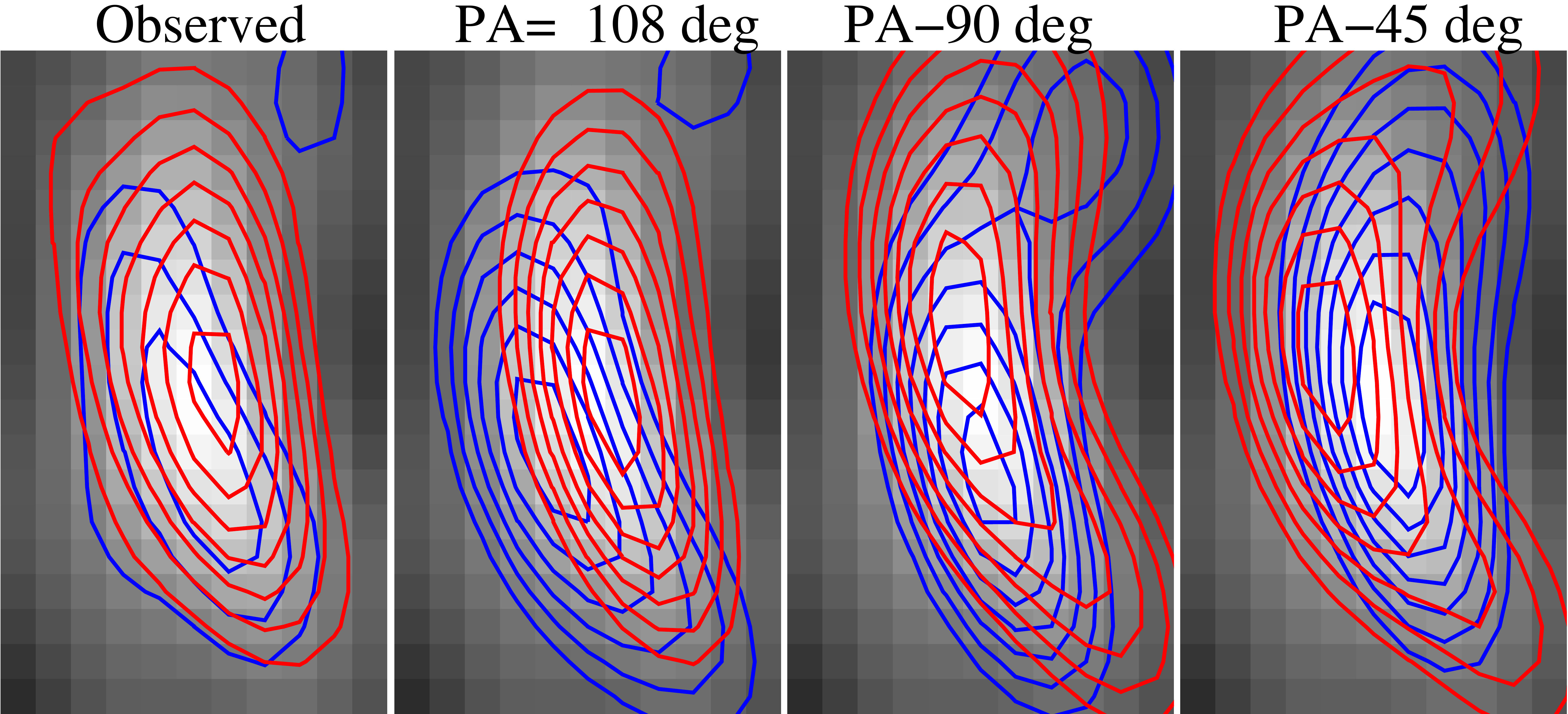}  
     }
   \caption{{\it Left panel:} observed velocity gradient. The grayscale image is a zoom into the CO(7--6) map shown in Fig.\,\ref{fig:obs}. The blue and red contours correspond to the CO(7--6) line emission integrated in the ranges $[-135,-15]$ and  $[-15,105]$\,km\,s$^{-1}$, respectively. The contours start at 3-$\sigma$ and are spaced by $\sigma$=104\,mJy\,beam$^{-1}$\,km\,s$^{-1}$. {\it 2nd to 4th panels:} Simulated velocity gradients with two lensed circular Gaussian sources of FWHM $\rho$=120\,mas, separated by 110\,mas and with a position angle of 108, 18 and 63\,$\deg$, east from north, respectively (see Section\,\ref{sec:lensmodel}).}
   \label{fig:velgrad}%
\end{figure}

\begin{figure}
   \centering
   
     \includegraphics[width=6cm]{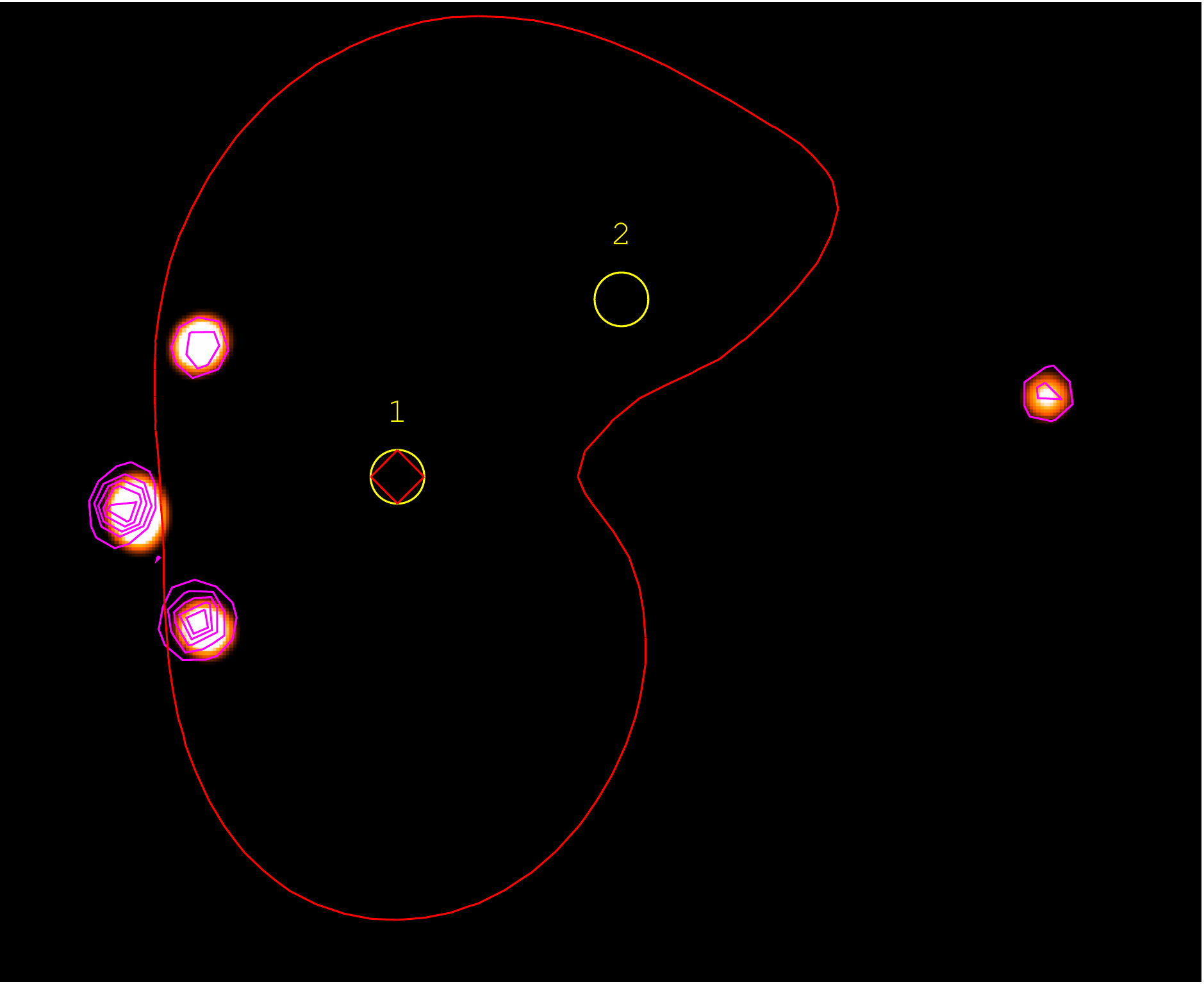}
   \caption{Four quasar images obtained with a lens model built following \citet{2000ApJ...544L..35K}. The HST image is overlaid in magenta contours. The red lines represent the critical lines and the two yellow circles show the positions of the two lensing galaxies.}
   \label{fig:lensmodel}%
\end{figure}

\begin{figure*}
   \resizebox{!}{5.6cm}{
   \includegraphics[height=5cm, trim=0 0 0 0,clip]{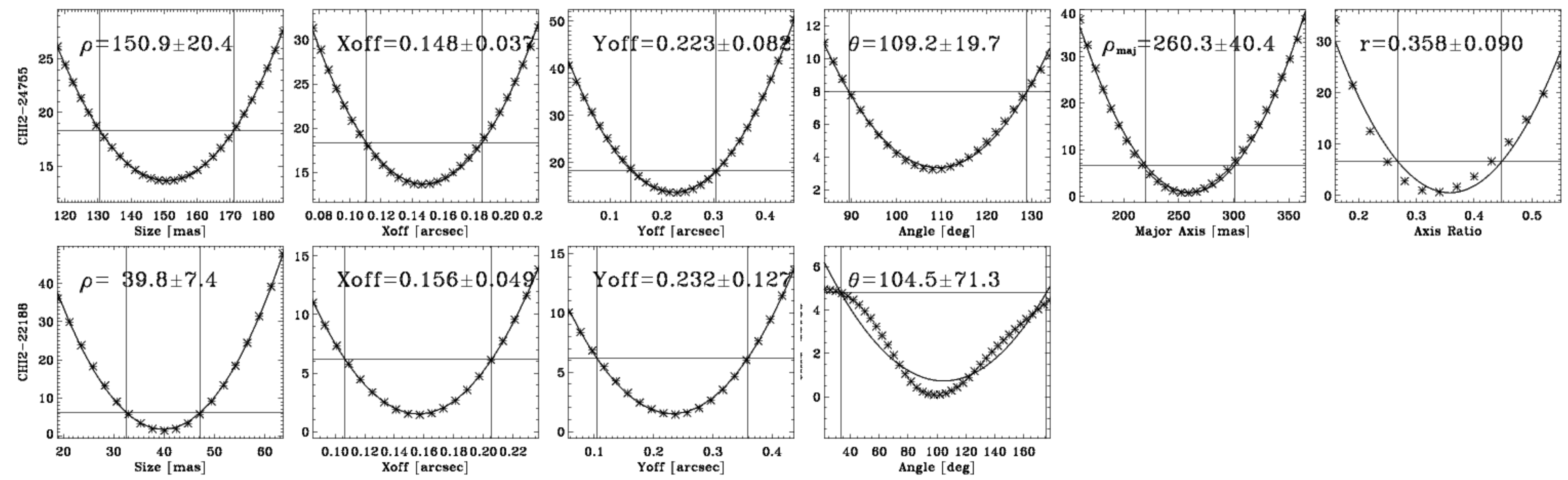}
   }
   \caption{$\chi^2$ variations with the source parameters. The upper panels show the variations of $\chi^2$-24755 for the line maps  and the lower panels show $\chi^2$-22188 for the continuum maps. From left to right the abscissae are the source size (FWHM), $\rho$,  assuming a circular Gaussian source; the offset in RA, $x_{\rm off}$; the offset in Dec, $y_{\rm off}$;  the position angle $\theta$ assuming an elliptical Gaussian source;  the source major axis, $\rho_{\rm maj}$, and the source axis ratio $r=\rho_{\rm min}/\rho_{\rm maj}$. { The horizontal line shows the 68\% confidence level ($1\times\sigma$) taking into account the number of free parameters.}}
   \label{fig:chi2}%
\end{figure*}

\begin{figure}
   \centering
   
   \includegraphics[angle=90, width=8cm]{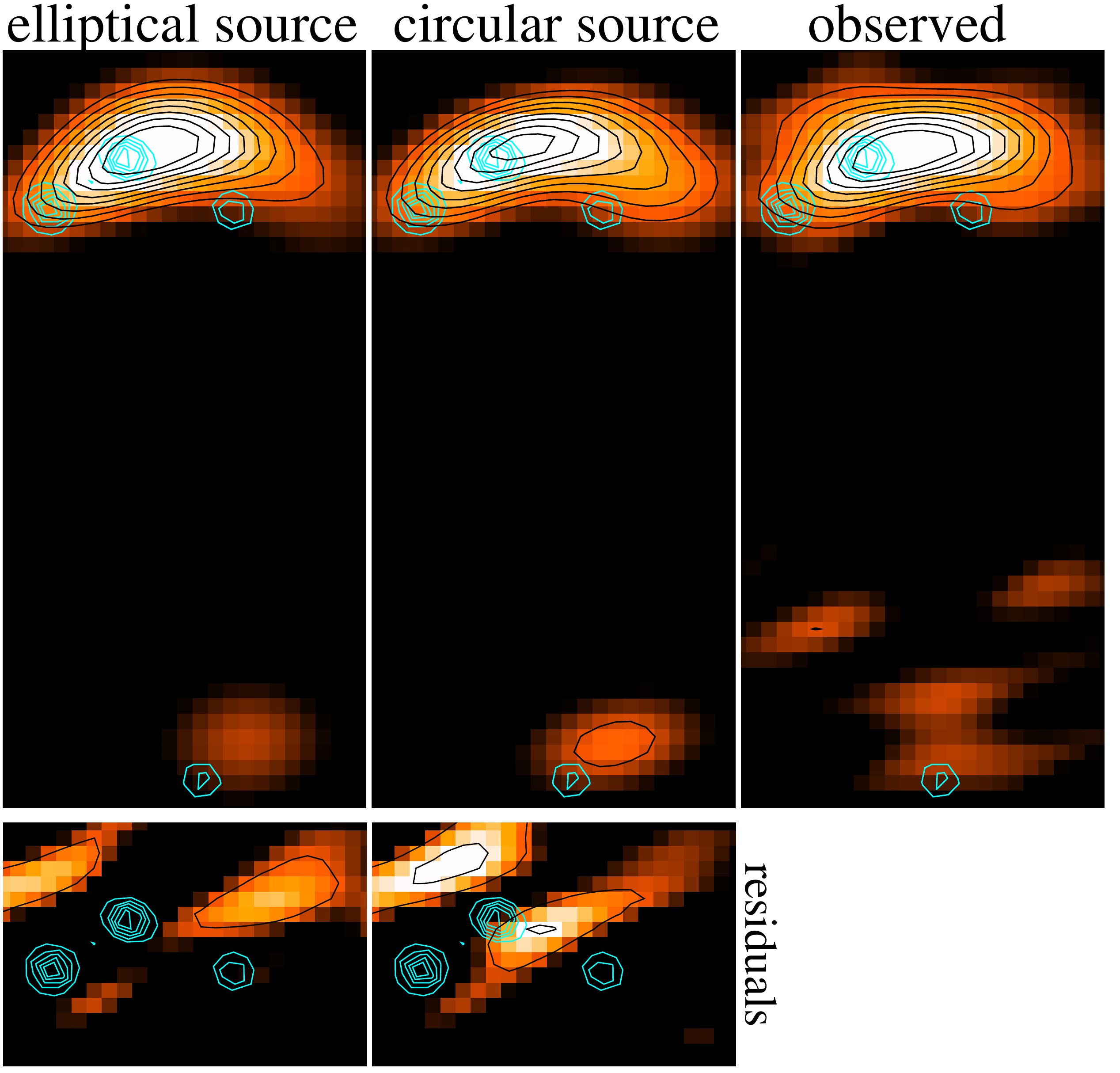}
     
\caption{{\it Upper panel:} CO(7--6) map shown in the Fig.\,\ref{fig:obs}. {\it Middle panel:} The best-fit Gaussian circular source with $\rho$=151\,mas lensed and convolved by the PdBI clean beam. {\it Lower panel:} 
The best-fit elliptical source lensed and convolved with the PdBI clean beam. The ellipse parameters are  $\rho_{\rm maj}$=260\,mas, $\rho_{\rm min}/\rho_{\rm maj}$=0.36, and $\theta$=108\,$\deg$. The left insets show the residuals at the A images, the contours are at 1$\times\sigma$ and 2$\times\sigma$.}
   \label{fig:lensfit}%
\end{figure}

\begin{figure}
\centering
\resizebox{7cm}{!}{
\includegraphics[height=5cm, trim=0 0 0 0,clip]{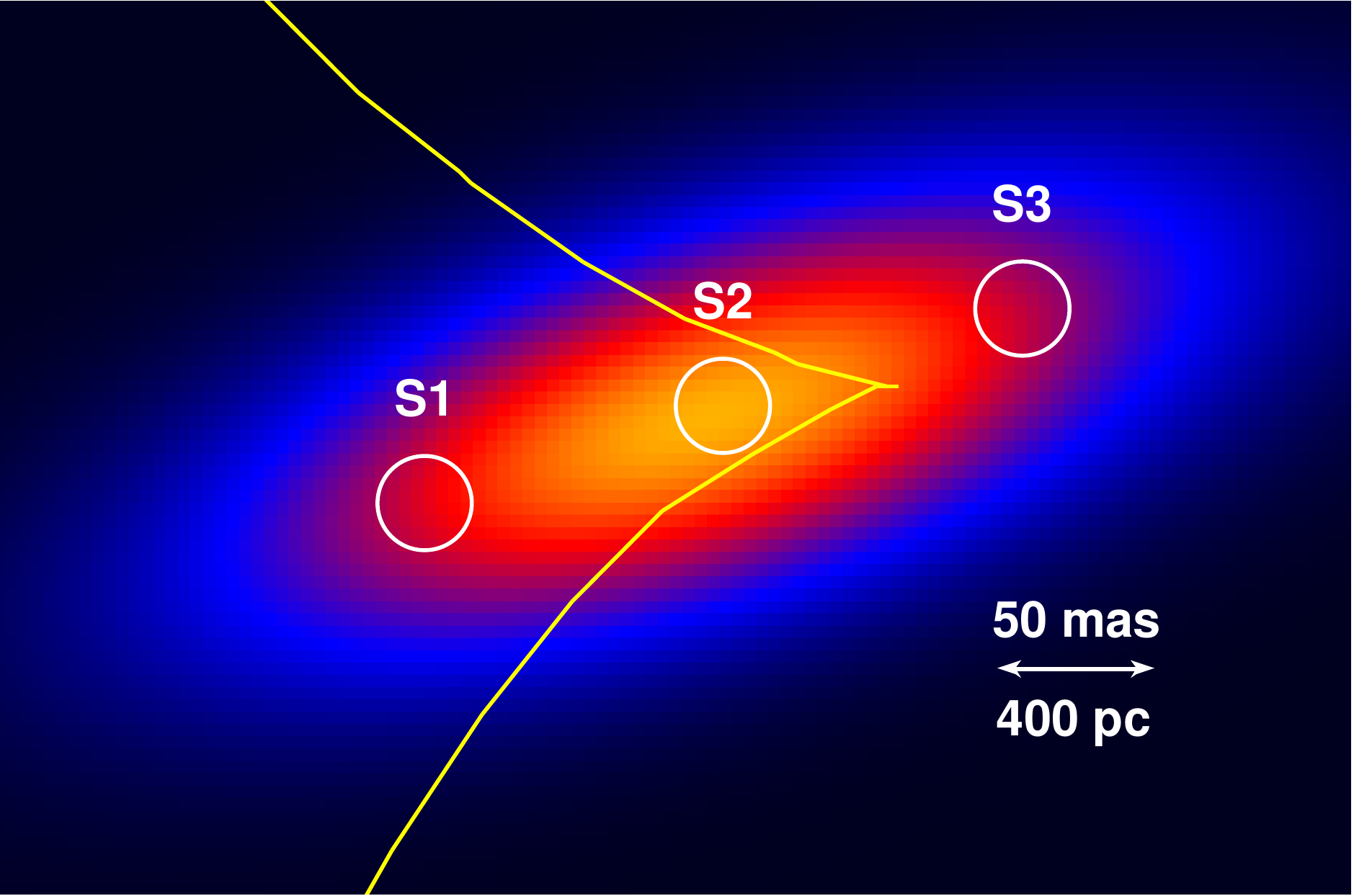}
}
\caption{Color image representing the best-fit elliptical Gaussian source to the CO(7--6) emission. The yellow curve shows the caustic line derived from the lens model. To quantify the effect of differential magnification we considered three virtual sources. S1, S2, and S3 are magnified by 18.8, 39.3, and 13.8, respectively.}
\label{fig:sourceplane}
\end{figure}

\subsection{Observations}

The data were collected with PdBI during $\sim$18 hours on-source in good observing conditions in February and March 2007. The receivers were tuned at 212.485\,GHz.
Pointing and focus measurements were performed at intervals shorter than one hour.  
Several quasars within 10\,$\deg$ from the source were used as calibrators: MCW349, 0851+202, 0906+015, 0923+392, and J0418+380. 
Data reduction was performed using the GILDAS standard package (\url{http://www.iram.fr/IRAMFR/GILDAS/}). The data were averaged in 46 velocity bins of 30 km\,s$^{-1}$ each. 
The dirty map was reconstructed on a square grid with a pixel size of 80 mas using natural weighting.  
The map was cleaned with the CLEAN method \citep{1974A&AS...15..417H}. The clean beam is an elliptic Gaussian with a full width at half maximum (FWHM) of 0.72$''$ and 0.27$''$ along the major and minor axes, respectively, and a position angle of 15\,$\deg$ from north to east. The noise level is 1.7\,mJy\,beam$^{-1}$ in each velocity channel.

\subsection{Flux, morphology, and velocity gradient}\label{sec:morpho}

The integrated CO(7--6) line map, the continuum map, and the integrated spectrum are shown in the Figure\,\ref{fig:obs}. The HST observations of the quasar with the astrometry calibrated to match that given by \citet{2000ApJ...544L..35K} are overlaid in contours.
The line width, FWHM=120$\pm$14\,km\,s$^{-1}$, the line peak, $S_{\rm max}$=48$\pm$5\,mJy, and the line redshift are consistent with those obtained by \citet{2012ApJ...753..102W} from independent PdBI observations. However, due to a higher noise level we do not detect the weak line wings reported by these authors that could be the signature of an outflow. The line intensity is $I_{\rm CO(7-6)}=6.1\pm 1.0$\,Jy\,km\,s$^{-1}$, which translates into a line luminosity  $L^{\prime}_{\rm CO(7-6)}=(4.36\pm 0.8)\times \mu^{-1} \times 10^{10}$\,K\,km\,s$^{-1}$\,pc$^{2}$, where $\mu$ is the lensing magnification factor. The continuum flux, $S_{\rm 212GHZ}=(7.4\pm 0.9)\times \mu^{-1}$\,mJy, is also consistent with other submm continuum measurements as shown Fig.\,\ref{fig:obs}. The best-fit modified black body spectral energy distribution (SED) with $\beta=1.5$ gives a dust temperature $T_{\rm d}=39\pm 3$\,K and a total FIR luminosity $L_{\rm FIR}=273\pm 45\times \mu^{-1} \times 10^{11}$\,L$_{\odot}$.

The line and continuum emissions show similar morphologies that are consistent with the configuration of the four quasar images observed with HST. { The B image is tentatively detected at the 2.6-$\sigma$ level, consistently with the B/A flux ratio of 0.21 measured by \citet{2012ApJ...753..102W}}. The overlay also suggests that the PdBI observations are offset with respect to HST observations, as was noted by \citet{2012ApJ...753..102W}.

Fitting four point sources (i.e., beams) with relative positions fixed to those observed with HST to the CO emission leads to large residuals. This suggests that the molecular gas around the quasar is resolved.
This hint is further supported by  the evidence of a velocity gradient across the map. Indeed, as shown in the Fig.\ref{fig:velgrad},  the emission corresponding to the blue half of the line is spatially offset from the emission corresponding to the red half of the line. The offset between the peaks of the blue and red emission is $\sim$0.3$''$ in the northwest direction, which is close to the beam width in this direction ($0.28''$). The offset is therefore significant and suggests a resolved dynamical structure.

\subsection{Source shape modeling}\label{sec:lensmodel}

To interpret the PdBI observations in terms of source properties we have built a lens model using the LENSTOOL software \citep{1996ApJ...471..643K,2007NJPh....9..447J,2009MNRAS.395.1319J} following the same approach as \citet{2000ApJ...544L..35K}. The model consists of three elliptical isothermal sphere  potentials corresponding to the nearby galaxy cluster, the main lensing galaxy, and a smaller neighboring galaxy. The four quasar images obtained are shown in the Fig.\,\ref{fig:lensmodel} with the HST observations overlaid in contours. In parallel and as a consistency check we compared our results with those obtained using a much simpler gravitational potential \citep{1998ApJ...501L...5B} and a different software \citep{hoai2013}.
%(Hoai et al.\ 2013).

To derive the source properties we proceeded by model-fitting in the Fourier plane. This method allows for a rigorous treatment of the errors and avoids artifacts inherent to imaging with interferometer data \citep[see e.g.][]{2012ApJ...756..134B}. For each source parameter value the sky image produced by LENSTOOL is Fourier-transformed and  sampled at the $(u, v)$ coordinates of the visibilites measured by PdBI.
The source parameters are varied to minimize  $\chi^2=\sum_i^N \|V^{\rm obs}_i-V^{\rm mod}_i\|^2/\sigma_i^2$, where $N=13710$ is the number of visibilities measured by PdBI, $V^{\rm obs}_i$ and $V^{\rm mod}_i$ are the measured and modeled complex visibilities, and $\sigma_i$ are the noise root mean squares (RMS) of the measurements. In addition to the source shape parameters we considered an astrometric offset in RA and Dec ($x_{\rm off}$ and $y_{\rm off}$) and a scaling factor ($f$, equivalent to a source flux parameter). { Any significant offset of the CO and/or the continuum emission from the QSO in the source plane was excluded because it would produce a different configuration for the  images. Such an offset is generally observed in merging systems, which does not seem to be the case here given the single narrow line. The number of free parameters was taken into account when computing the uncertainties following \citet{2002nrc..book.....P}}.  

We first used a simple circular Gaussian source model for which the only shape parameter is $\rho$, the FWHM of the Gaussian. The best-fit values obtained are $\rho=151\pm 20$ and $40\pm 8$\,mas for the line and continuum emission, respectively.  
The offsets obtained are $x_{\rm off}=0.15\pm 0.04$ and $y_{\rm off}=0.22\pm 0.08''$, consistent with the uncertainties in the astrometry of the HST map. The corresponding $\chi^2$ curves and best-fit sky images are shown in the Figs.\,\ref{fig:chi2} and \ref{fig:lensfit}. The offsets obtained in the other fits described below are similar within the uncertainties.
These results show that the region emitting the CO(7--6) line is clearly resolved and that the region emitting the continuum is resolved with a confidence of $\sim5\sigma$.

A galaxy is not expected to be circular on the sky. However, fitting a non-circular model would require fitting two more parameters (elongation and orientation), which seems too much for the signal-to-noise ratio and the resolution of our data. We therefore followed a  two-step approach. We first fitted the orientation of an elliptical Gaussian assuming a fixed and arbitrary elongation for the source, namely an elliptical Gaussian source with an axis ratio of $\rho_{\rm min}/\rho_{\rm maj}=0.5$. To keep the size of the source consistent with $\rho$, the FWHM  obtained with the circular source fit, we derived $\rho_{\rm min}$ and $\rho_{\rm maj}$ from $(\rho_{\rm min}^2+\rho_{\rm maj}^2/2)^{1/2}=\rho$. We thus obtained a position angle of $\theta=108\pm 20$ and $104\pm 71\,\deg$ from north to east for the line and the continuum, respectively (Fig.\,\ref{fig:chi2}). It is remarkable that for the line emission the minimum $\chi^2$ reached is significantly lower than for the circular source best-fit {\citep[$\Delta\chi^2$=10, i.e., a significance of 95.5$\%$ for four free parameters,][] {2002nrc..book.....P}},  indicating that the source is indeed elongated.  This is not the case for the continuum, and the large uncertainty on $\theta$ shows that the position angle of the continuum source cannot be constrained by our data. We checked that changing the source axis ratio does not affect the value of $\theta$ obtained for the line emission within the uncertainties. 

In a second step we fixed the position angle to $\theta=108\,\deg$ and we simultaneously fitted  the major axis FWHM of the elliptical Gaussian, $\rho_{\rm maj}$ and its axis ratio, $r=\rho_{\rm min}/\rho_{\rm maj}$. This was done for the line-integrated map only because the continuum does not appear to be elongated. The results are $\rho_{\rm maj}=260\pm 40$\,mas and $r=0.36\pm 0.09$,  
excluding a circular source with a confidence $>7\sigma$.
The improvement with respect to the circular source is also noticeable in the sky image (Fig.\,\ref{fig:lensfit}). We checked that these results are robust against small variations of the lens model parameters within the uncertainties.

If this is a rotating disk, there should be a velocity gradient along the major axis. To test whether this solution is consistent with the velocity gradient noticed in Section\,\ref{sec:morpho}, we replaced the elliptical model by two circular Gaussian sources of FWHM, $\rho=120$\,mas, separated by 110\,mas along the major axis and assumed that the northern source is red and the other is blue. The result is shown in the second panel of Fig.\,\ref{fig:velgrad}. As can be seen, the velocity gradient thus produced looks similar to the observed one. This would not be the case for other orientations, as illustrated in the two other panels. This good agreement with the kinematics information supports our fit results.

\section{Discussion}

From our best-fit models we derived global magnification factors of $\mu\!\!=$18.1$\pm$3.6 and 37.7$\pm$7.2 for the line and the continuum respectively, yielding $L^{\prime}_{\rm CO(7-6)}=(2.42\pm 0.50) \times 10^{9}$\,K\,km\,s$^{-1}$\,pc$^{2}$ and $S_{\rm 212GHZ}=(196\pm 30)$\,$\mu$Jy. 

{ Assuming $\mu$=22,  \citet{2011ApJ...739L..32R} measured $L^{\prime}_{\rm CO(1-0)}\!=\!3.39\!\pm\!0.48$\,K\,km\,s$^{-1}$\,pc$^{2}$,  which implies $L^{\prime}_{\rm CO(1-0)}/L^{\prime}_{\rm CO(7-6)}=1.2$ if we assume $\mu$=18.1 for both CO lines. These authors also noted that the corresponding gas mass $M_{\rm gas}\!=\!(2.3\pm 0.5)\times 10^{9}$\,M$_{\odot}$ \citep[using the conversion factor $\alpha=0.8$ from][]{1998ApJ...507..615D} is at the low end of the observed range of gas masses in QSOs.}

Assuming that the submm continuum measurements are all emitted from the same region (no differential magnification) and that they can be modeled with a single modified black body SED as in Fig.\,\ref{fig:obs}, the total FIR luminosity is $L_{\rm FIR}=(7.2\pm 1.5) \times 10^{11}$\,L$_{\odot}$. If the FIR is entirely produced by star formation, it corresponds to an SFR=110\,M$_{\odot}$\,yr$^{-1}$ \citep{1998ApJ...498..541K, 1998ARA&A..36..189K}.
 { From the best-fit SED and making the same assumptions as \citet[][]{2007A&A...475..513B}, we derive a dust mass $M_{\rm d}\!\!=\!\!(5.9\pm 1.2)\times 10^7$\,M$_{\odot}$, corresponding to a gas-to-dust ratio $M_{\rm gas}/M_{\rm d}\!\!=\!\!45\pm 17$, i.e., similar to that measured in the $z$=6.4 QSO { SDSS\,J1148+5251} \citep{2012MNRAS.427L..60V}.

The FIR and CO luminosities are compatible with the $L_{\rm FIR}-L^{\prime}_{\rm CO}$ correlation measured in quasars from low to high redshifts by \citet{2011ApJ...730..108R} and do not match the correlation found by \citet{2013MNRAS.429.3047B} for SMGs, $L^{\prime}_{\rm CO}$ being a factor $\sim 3$ too low. RXJ0911's host properties would therefore be consistent with the transition scenario from SMGs to QSOs.

  Ignoring the differential magnification, i.e., using the same magnification factor for both the CO and the FIR emission, would have lead to an overestimate of the star formation efficiency measured by $L_{\rm FIR}/L^{\prime}_{\rm CO}$.
 This case clearly illustrates why this effect needs to be taken into account, in particular in QSOs where the FIR may come from a more compact region than the CO lines. The star formation efficiency may thus have been overestimated in several unresolved lensed QSO hosts.}

The best-fit source shape parameters can be used to obtain a tentative estimate of the geometry.
At $z\!\!=\!\!2.796$, 1$''$ corresponds to 7.98\,kpc, the characteristic radius of the CO disk is therefore $R\!=\!1\pm 0.2$\,kpc, and its inclination angle is $i=90-\arcsin(r)=69\pm 6\deg$. This high inclination seems to contradict the remarkably narrow line observed, which suggests a face-on disk. However, the corresponding dynamical mass, $M_{\rm dyn}=R\times \Delta v^2\times G^{-1} \times \sin i^{-2}=(3.9\pm 0.9)\times 10^9$\,M$_{\odot}$, is a factor 2 larger than the molecular gas mass, implying a gas fraction of 50\%, similar to the one observed in other high-z QSOs \citep[e.g.,][]{2009ApJ...690..463R,2005ARA&A..43..677S}. In other words, the line may be truly narrow { (the line width corrected for inclination would be $W=129\pm 20$\,km\,s$^{-1}$)} because this quasar host may lie on the 'light' end of the mass distribution of quasar hosts \citep[][]{2006AJ....131.2763C}. This interpretation is also consistent with the X-ray luminosity, $L_{\rm X}=1.4\times 10^{44}$\,erg\,s$^{-1}$ \citep[][assuming the same magnification factor as for the FIR, { i.e., 37.7}]{2009ApJ...690.1006F} because according to the well-known $M_{\rm BH}-\sigma_{\star}$ relation in local elliptical galaxies \citep{2009ApJ...698..198G} and following \citet{2013MNRAS.429.3047B}, the line width expected is $W\sim\sigma_{\star}/1.5\sim 135$\,km\,s$^{-1}$ { for $M_{\rm BH}=10^{7.7}$M$_{\odot}$}.  Thus, RXJ0911 appears to be a scaled-down version of the QSOs usually found at high-$z$, residing in a lighter halo, containing less gas, forming fewer stars and having a lighter central black hole.

However, the line may still be broader than observed due to the effect of differential magnification. Indeed, as was recently noted by \citet{hoai2013}, the source is near the caustic cusp and differential magnification is expected to be significant. In fact, as shown in Fig.\,\ref{fig:sourceplane}, the caustic lines cross our best-fit source model mainly along the minor axis where the line-of-sight velocities should be close to the systemic velocity. Differential magnification should therefore boost the center of the spectral line when integrated over the entire source, resulting in a narrower profile. To illustrate this effect we computed the magnification factor of three virtual circular Gaussian sources separated by 100\,mas and located along the major axis, as shown in Fig.\,\ref{fig:sourceplane}. We find that the central source S2 is more magnified than S1 and S3, by a factor 2.1 and 2.8, respectively.  Correcting the line profile for this effect would require  estimating the rotation curve from higher resolution observations. { A higher sensitivity is also required to investigate whether the broad line wings discovered by \cite{2012ApJ...753..102W} could be caused by this effect. In a similar study of another lensed quasar, \citet{2008ApJ...686..851R} found a much weaker differential effect on the CO line. This is because the CO region observed in their source is much more extended than the region delimited by the caustic lines; this is not the case in RXJ0911}.

This case study shows how gravitational lensing can provide insights into sources that are scaled-down versions of the usually selected ones. But it also underlines the possible bias introduced by lensing on the integrated measurements due to differential magnification. High-resolution observations are required to correct for this effect and to access the physical properties of the sources. ALMA will probably allow us to study the molecular gas morphology and kinematics in this and other lensed QSOs and SMGs in more detail. { In particular, ALMA will be essential for studying the scaling relations toward the lower intrinsic gas masses. Indeed, the few other lensed quasars with similar low molecular gas masses lack good lens models \citep[][and references therein]{2011ApJ...730..108R}}. This will allow us to better understand the  relation between molecular gas mass, star formation rate, and  total mass, and to identify possible differences between the properties of QSO hosts and SMGs in a more robust way.

\bibliographystyle{aa}
\bibliography{rxj09}

\begin{acknowledgements}
    We thank the IRAM staff for carrying out these PdBI observations and for helping us with the data reduction. We are very grateful to the referee for the insightful and constructive suggestions that helped to improve the paper. We would like to acknowledge the financial support from the Vietnamese National Foundation for Science and Technology Development.
\end{acknowledgements}

\end{document}